\documentclass[pre,onecolumn]{revtex4}
\usepackage{epsfig}
\normalsize

\begin{document}

\title{LOCAL QUASI-EQUILIBRIUM DESCRIPTION OF SLOW RELAXATION SYSTEMS}

\author{I. Santamar\'{\i}a-Holek, A. P\'{e}rez-Madrid and J. M. Rub\'{\i}}

\affiliation{Department de F\'{\i}sica Fonamental\\
 Facultat de F\'{\i}sica\\
 Universitat de Barcelona\\
 Diagonal 647, 08028 Barcelona, Spain\\
 }

\begin{abstract}
We present a dynamical description of slow relaxation processes based on the
extension of Onsager's fluctuation theory to systems in local quasi-equilibrium.
A non-Markovian Fokker-Planck equation for the conditional probability density
is derived and from it, we obtain the relaxation equation for the moments. We
show that the fluctuation-dissipation theorem can be formulated in terms of
the temperature of the system at local quasi-equilibrium which is related to
that of the bath by means of a scaling factor \textbf{}revealing lack of thermal
equilibrium. Our theory may be applied to a wide variety of systems undergoing
slow relaxation. We discuss in particular slow dynamics in glassy systems and
Brownian motion in a granular gas.
\end{abstract}
\maketitle

\section{Introduction}

Systems whose global evolution results from the existence of a wide variety
of time and length scales frequently exhibit slow relaxation dynamics, which
is manifested through the aging behaviour of the correlations and the nonexponential
decay of the response function \cite{angell}-\cite{stillinger}. \textbf{}To
characterize the dynamics of these systems is one of the most challenging problems
of nonequilibrium statistical mechanics nowadays. Many efforts have been made
in this sense by proposing different approaches including spin models, stochastic
dynamics and thermodynamical approaches (see, for example, Refs. \cite{Anderson}-\cite{theo}).

Experiments performed in diverse systems as in amorphous polymers \cite{struick}
and supercooled liquids and glasses \cite{angell}, \cite{israeloff}-\cite{coll-glasses},
seem to indicate that these systems undergo, in general, non-Markovian and non-stationary
dynamics leading to the peculiar behaviour of the two-time correlation functions
and to \textbf{}the dependence of the relaxation times on the initial time. 

In this paper we present a dynamical description of slow relaxation systems
based on the formulation of a Fokker-Planck equation for non-Markovian processes.
This equation is obtained through the generalization of the Onsager theory,
originally proposed to account for the dynamics of fluctuations around equilibrium
states \cite{onsager} and extended to nonequilibrium stationary states \cite{dorfman},
to nonequilibrium \emph{aging} states.

An important feature of the dynamics of these systems is that at the time scales
of experiments, they still relax and never reach thermal equilibrium with the
bath. We have assumed the existence \textbf{}of a local quasi-equilibrium state
\cite{virasoro} \textbf{}characterized by a non-stationary probability density
which can be used to determine average \textbf{}quantities as, for example,
the local quasi-equilibrium temperature \cite{librozwanzig}. The entropy of
the system has been expressed in terms of the conditional probability density
by means of the Gibbs entropy postulate, and used to derive the generalized
Fokker-Planck equation by taking into account probability conservation and the
rules of nonequilibrium thermodynamics \cite{degroot}. \textbf{}For stationary
processes, this equation \textbf{}was derived in \cite{adelman} and \cite{pop-mori},
and has subsequently been treated in the literature of stochastic processes
in \cite{oxtoby}-\cite{FP-nonmarkovian}. 

The unified description of slow relaxation systems we propose can be applied
to, among others, glassy systems \cite{kurchan}, granular matter \cite{dufty1}
and anomalous diffusion problems \cite{tokuyama}, \cite{oliveira}.

The paper is distributed as follows. In Sec. \textbf{II}, we will analyze the
non-stationary dynamics of slow relaxation systems under the framework of a
generalized Onsager theory. Section \textbf{III} is devoted to the formulation
of relaxation equations and of the generalized fluctuation-dissipation theorem
for non-stationary systems. An illustration of the theory presented is given
in Sec. \textbf{IV} by analyzing the dynamics of systems having two well-separated
time scales (glassy systems). In Sec. \textbf{V}, we will discuss the Brownian
motion in a granular gas. We will compare our results with those obtained by
means of kinetic theory and simulations. In the discussion section we will \textbf{}summarize
our main results.

\section{Fokker-Planck equation for non-stationary dynamics }

The non-stationary dynamics of slow relaxation systems can be characterized
at the mesoscopic level by means of the conditional probability density \( P(\underline{\alpha }_{0},t_{0}|\underline{\alpha },t) \),
which satisfies the normalization condition \( \int P(\underline{\alpha }_{0},t_{0}|\underline{\alpha },t)d\underline{\alpha }=1 \).
This quantity depends in general on a set of fluctuating variables \( \underline{\alpha }=(\alpha _{1},...,\alpha _{n}) \),
or fields \( \underline{\alpha }(\vec{k}) \), \cite{hydrofluc}, necessary
to univocally determine the state of the system during the relaxation and which
may represent, for instance, fluctuating positions and velocities, order parameters,
\textbf{}reaction coordinates or hydrodynamic fields. \textbf{}

The evolution in time of \( P(\underline{\alpha }_{0},t_{0}|\underline{\alpha },t) \),
with \( \underline{\alpha }_{0} \) the state of the system when the fluctuation
takes place at \( t_{0} \) (in \textbf{}glassy systems literature it is referred
to as waiting time or time elapsed after quenching)\textbf{,} is governed by
the continuity equation 
\begin{equation}
\label{continuidad alfa}
\frac{\partial }{\partial t}P(\underline{\alpha }_{0},t_{0}|\underline{\alpha },t)=-\frac{\partial }{\partial \underline{\alpha }}\cdot \left[ P(\underline{\alpha }_{0},t_{0}|\underline{\alpha },t)\underline{v}_{\underline{\alpha }}(\underline{\alpha }_{0},\underline{\alpha };t_{0},t)\right] ,
\end{equation}
 expressing probability conservation during its natural evolution in \( \underline{\alpha } \)-space.
The probability current \( P(\underline{\alpha }_{0},t_{0}|\underline{\alpha },t)\underline{v}_{\underline{\alpha }}(\underline{\alpha }_{0},\underline{\alpha };t_{0},t) \),
with \( \underline{v}_{\underline{\alpha }}(\underline{\alpha }_{0},\underline{\alpha };t_{0},t) \)
a stream velocity in \( \underline{\alpha } \)-space, has to be determined
in order to render the dynamical description complete. \textbf{}

This task can be accomplished by assuming that the scheme of nonequilibrium
thermodynamics \cite{degroot} remains valid in \( \underline{\alpha } \)-space.
Thus, a linear relationship can be established between the diffusion current
\( P(t_{0}|t)\underline{v}_{\underline{\alpha }}(t_{0},t) \) and its conjugate
force in \( \underline{\alpha } \)-space, which can be obtained from the entropy
production calculated in the framework of Onsager's theory \cite{degroot}.
To derive the entropy production we will assume that, for sufficiently long
times \( t \), the system reaches a local quasi-equilibrium state, characterized
by the entropy per unit mass \( s_{qe}(t) \) and the probability density \( P_{qe}(\underline{\alpha },t) \),
related formally to the conditional probability through \( \lim _{t\rightarrow \infty }P(\underline{\alpha }_{0},t_{0}|\underline{\alpha },t)=P_{qe}(\underline{\alpha },t) \). 

In accordance with \cite{inertial}, the variation of the entropy \textbf{}\( s_{qe}(t) \)
satisfies the Gibbs equation

\begin{equation}
\label{Gibbs ec}
T(t)\delta s_{qe}(t)=\delta e_{qe}(t)-\int \mu _{qe}(t)\delta P(t)d\underline{\alpha },
\end{equation}
 where the temperature \( T(t) \) \textbf{}of the system at local quasi-equilibrium
will be assumed to be only a function of time, \( e_{qe}(t) \) is the mean
internal energy, and \( \mu _{qe}(t) \) the chemical potential at local quasi-equilibrium,
which do not depend on the \( \underline{\alpha } \)'s. \textbf{}For simplicity
we have omitted the explicit dependence of the probability density on phase
variables. The entropy production \( \sigma  \) corresponding to the diffusion
process in \( \underline{\alpha } \)-space can be computed by using the Gibbs
entropy postulate \cite{kampenslow}

\begin{equation}
\label{p. gibbs}
s(t)=-\frac{k_{B}}{m}\int P(t_{0}|t)\ln \frac{P(t_{0}|t)}{P_{qe}(t)}d\underline{\alpha }+s_{qe}(t),
\end{equation}
 where \( s(t) \) is the entropy of the system, \( k_{B} \) the Boltzmann's
constant and \( m \) the molecular mass. Notice that, according with mesoscopic
nonequilibrium thermodynamics (MNET), the entropy \( s_{qe}(t) \) appearing
in Eq. (\ref{p. gibbs}), constitutes a macroscopic quantity whose variations
are governed by Eq. (\ref{Gibbs ec}). The evolution in time of \( s_{qe}(t) \)
can also be described with the MNET scheme by using the expression of Eq. (\ref{Gibbs ec}),
and the corresponding Eq. (\ref{p. gibbs}) referred to the equilibrium state
(see, for example, \cite{inertial} and \cite{MNETmazur}). In that case, by
taking the variation \( \delta s_{qe}=s_{qe}(\left\{ P_{qe}+\delta P\right\} )-s_{qe}(\left\{ P_{qe}\right\} ) \)
at first order in \( \delta P \) and combining the resulting expression with
Eq. (\ref{Gibbs ec}), we obtain 

\textbf{
\begin{equation}
\label{P0 a la entropic}
P_{qe}(t)=P_{e}\, e^{\frac{m}{k_{B}\, T(t)}[\mu _{qe}-e_{qe}]},
\end{equation}
} which incorporates the topology of the energy landscape through \( e_{qe} \),
and where \( P_{e} \) is the equilibrium probability distribution. \textbf{}

By combining the rate of change of \( s_{qe}(t) \) obtained from Eq. (\ref{Gibbs ec})
with the time derivative of Eq. (\ref{p. gibbs}), after using Eq. (\ref{continuidad alfa})
and integrating by parts assuming that \textbf{\( P(t_{0}|t)\underline{v}_{\underline{\alpha }}(t_{0},t) \)}
vanishes at the boundary, we obtain \cite{gradtemp}

\begin{equation}
\label{produccion S}
\frac{\partial }{\partial t}s(t)=\frac{1}{T(t)}\frac{\partial }{\partial t}e(t)-\frac{1}{T(t)}\int P(t_{0}|t)\underline{v}_{\underline{\alpha }}(t_{0};t)\cdot \frac{\partial }{\partial \underline{\alpha }}\left[ \frac{k_{B}T(t)}{m}\ln \frac{P(t_{0}|t)}{P_{qe}(t)}+\mu _{qe}(t)\right] d\underline{\alpha },
\end{equation}
 where \( \delta e(t)=\delta e_{qe}(t)+\frac{k_{B}T(t)}{m}\int P(t_{0}|t)\delta \ln P_{qe}(t)d\underline{\alpha } \)
is the variation of the nonequilibrium \textbf{}internal energy \( e(t) \).
If no external thermodynamic forces as, for example, velocity or temperature
gradients act on the system, it may be shown by using the Gibbs-Duhem relation
that \( \delta e(t)=0 \). As a consequence of this, the entropy production
is only due to the underlying diffusion process in \( \underline{\alpha } \)-space.
Moreover, in the last term of Eq. (\ref{produccion S}) one may identify the
nonequilibrium ``chemical potential'' \( \mu (t_{0};t) \) as 

\begin{equation}
\label{potencial quimico}
\mu (t_{0};t)=\frac{k_{B}T(t)}{m}\ln \frac{P(t_{0}|t)}{P_{qe}(t)}+\mu _{qe}(t).
\end{equation}
 Important to notice is the fact that \( \mu (t_{0};t) \) may depend, in general,
on interaction potentials \textbf{}or \textbf{}entropic barriers \textbf{}\cite{Entropic barriers}
through the corresponding dependence of \( P_{qe}(t) \).

According to the rules of nonequilibrium thermodynamics expressing currents
as linear functions of the conjugate thermodynamic forces \cite{degroot}, from
the entropy production (\ref{produccion S}\textbf{)} the velocity \textbf{\( \underline{v}_{\underline{\alpha }}(t_{0};t) \)}
can now be expressed in terms of \( \frac{\partial }{\partial \underline{\alpha }}\mu (t) \),
obtaining 

\begin{equation}
\label{linear laws 2}
\underline{v}_{\underline{\alpha }}(t_{0};t)=-\underline{\underline{B}}(t_{0};t)\cdot \left[ \underline{X}(\underline{\alpha },t)+\frac{k_{B}T(t)}{m}\frac{\partial }{\partial \underline{\alpha }}\ln P(t_{0}|t)\right] ,
\end{equation}
where we have identified the generalized force \cite{FP-nonmarkovian}, \cite{degroot}

\begin{equation}
\label{X force}
\underline{X}(\underline{\alpha },t)=-\frac{\partial }{\partial \underline{\alpha }}\left[ \frac{k_{B}T(t)}{m}\ln P_{qe}(t)\right] ,
\end{equation}
and introduced the coefficients \( \underline{\underline{B}}(t_{0};t) \) which
may, in general, depend on the state variables \textbf{\( \underline{\alpha } \)}.
\textbf{}These coefficients, incorporating memory effects through their time
dependence, can be determined in terms of the conditional relaxation functions
of the corresponding variables \( \underline{\alpha } \), \cite{degroot}.
Notice \textbf{}that, in general, the \( B_{ij} \)'s do not satisfy reciprocal
relations since for non-stationary systems and non-conservative interactions,
the time reversal symmetry and the principle \textbf{}of detailed balance are
not necessarily fulfilled. \textbf{}The time dependence of the \( B_{ij} \)'s
is the origin of anomalous diffusion, which is characterized by the power law
behaviour in time of the correlation function \cite{tokuyama}, \cite{oliveira},
and the nonexponential time decay of the relaxation functions \cite{angell}.

By substituting the expression of the probability current obtained from Eq.
(\ref{linear laws 2}) into Eq. (\ref{continuidad alfa}), we obtain the Fokker-Planck
equation

\begin{equation}
\label{fokker-planck general}
\frac{\partial P(t_{0}|t)}{\partial t}=\frac{\partial }{\partial \underline{\alpha }}\cdot \underline{\underline{B}}(t_{0};t)\cdot \left[ \underline{X}(\underline{\alpha },t)P(t_{0}|t)+\frac{k_{B}T(t)}{m}\frac{\partial P(t_{0}|t)}{\partial \underline{\alpha }}\right] .
\end{equation}
 Similar equations describing stationary non-Markovian dynamics were obtained
\textbf{}in Refs. \cite{adelman}-\cite{FP-nonmarkovian}, to study linear and
nonlinear transport. From Eq. (\ref{fokker-planck general}), it follows that
the characteristic relaxation times of the system \( B^{-1}_{ij}(t_{0};t) \),
are functions of the waiting time \( t_{0} \) as a consequence of the non-Markovian
nature of the dynamics. 

At the time scales considered, the system relaxes towards the time-dependent
local quasi-equilibrium state and then thermal equilibrium between system and
bath can not be established. The complete description of the dynamics can then
be performed by giving the relation between the system and bath temperatures.
In order to derive it, the influence of the bath is incorporated through the
second law \textbf{}\( \delta s_{T}=\delta s_{qe}+\delta s_{B}\geq 0 \), with
\textbf{\( s_{T} \)} the total entropy of the system and \( s_{B} \) that
of the bath. From this expression, by \textbf{}assuming that the variation \textbf{\( \delta s_{T} \)}
takes place at constant volume (see Eq. (\ref{Gibbs ec})), it can be expressed
as \( \delta s_{T}=[1-c]\frac{1}{T(t)}[\delta e_{qe}(t)-\delta g_{qe}(t)] \),
where we have taken into account the Gibbs equation (\ref{Gibbs ec}), used
the Gibbs free energy per mass unit \( \delta g_{qe}(t)=\int \mu _{qe}(t)\delta P(t)d\underline{\alpha } \)
and defined \( c=-\frac{\sigma _{B}}{\sigma }\geq 1 \), with \( \sigma _{B} \)
the entropy production of the bath. The equality occurs when the process is
reversible, i.e., \( \sigma _{B}=-\sigma  \). In a similar way, by using the
Gibbs equation of the bath we obtain \textbf{}\( \delta s_{T}=[1-c^{-1}]\frac{1}{T_{B}(t)}\delta e_{B}(t) \),
with \( e_{B}(t) \) and \( T_{B}(t) \) the energy and temperature of the bath.
A simple comparison of the expressions for \( \delta s_{T} \) leads to the
relation 

\textbf{
\begin{equation}
\label{Tfuncion deATo}
T(t)=AT_{B}(t),
\end{equation}
}where we have defined \( A=c\left[ \frac{\delta g_{qe}(t)-\delta e_{qe}(t)}{\delta e_{B\textrm{ }}}\right]  \).
Since \( \delta g_{qe}(t)=s_{qe}\delta T(t)-\mu _{qe}\delta \rho  \), we conclude
that the factor \( A \) also depends on the density and the temperature of
the system. \textbf{}When the process is reversible, \( A=1 \) since thermal
equilibrium between system and bath is attained. In this case, \( P_{qe}(t) \)
\textbf{}reduces to a local equilibrium \textbf{}probability density in which
the temperature is that of the bath \cite{Entropic barriers}, \cite{landau}.

\section{Relaxation equations and the fluctuation- dissipation theorem }

Once the Fokker-Planck equation has been \textbf{}derived, we will proceed to
establish the relaxation equations for the moments of the probability density.
The analysis of the general \textbf{}case can be accomplished by considering
that \textbf{}the coefficients are functions of the state variables \cite{FP-nonmarkovian}.
\textbf{}For the sake of simplicity, we will focus on the case in which the
coefficients \( B_{ij} \) are only functions of time. 

First, we will consider the two-time correlation function

\begin{equation}
\label{correlacion}
C_{\alpha _{i}\alpha _{0j}}(t_{0},t)\equiv \int \alpha _{i}\alpha _{0j}P(\underline{\alpha }_{0},t_{0})P(\underline{\alpha }_{0},t_{0}|\underline{\alpha },t)d\underline{\alpha }_{0}d\underline{\alpha },
\end{equation}
 whose evolution equation may be found by taking the time derivative of Eq.
(\ref{correlacion}), substituting Eq. (\ref{fokker-planck general}) and integrating
by parts, obtaining 

\begin{equation}
\label{evol. Caa}
\frac{d}{dt}C_{\alpha _{i}\alpha _{0j}}(t_{0},t)=-B_{il}(t_{0};t)\left\langle X_{l}(t)\alpha _{0j}(t_{0})\right\rangle ,
\end{equation}
where we have defined \( \left\langle X_{i}(t)\alpha _{0j}(t_{0})\right\rangle \equiv \int X_{i}(\underline{\alpha },t)\alpha _{0j}P(\underline{\alpha }_{0},t_{0})P(\underline{\alpha }_{0},t_{0}|\underline{\alpha },t)d\underline{\alpha }_{0}d\underline{\alpha } \).
This equation implicitly provides an operational definition of the transport
coefficients \( B_{ij} \), showing aging effects through their dependence on
the waiting time. The relation between the dissipation of the energy and the
response of the system to an external time-dependent force \( \underline{\epsilon }(t)=\underline{\epsilon }\Theta (t-t_{0}) \),
with \( \Theta (x) \) being the step function, may be determined by means of
the evolution equation for the first moment of the probability density, \( \left\langle \alpha _{i}(t)\right\rangle =\int \alpha _{i}P(\underline{\alpha }_{0},t_{0})P(\underline{\alpha }_{0},t_{0}|\underline{\alpha },t)d\underline{\alpha }_{0}d\underline{\alpha } \).
In the presence of the external force, the system is described by means of the
corresponding Fokker-Planck equation

\begin{eqnarray}
\frac{\partial P(t_{0}|t)}{\partial t}=\frac{\partial }{\partial \underline{\alpha }}\cdot \underline{\underline{B}}(t_{0};t)\cdot \left[ \underline{X}(\underline{\alpha },t)P(t_{0}|t)+\frac{k_{B}T_{B}(t)}{m}A\frac{\partial P(t_{0}|t)}{\partial \underline{\alpha }}\right] - &  & \nonumber \\
-\frac{\partial }{\partial \underline{\alpha }}\cdot \left[ \underline{\epsilon }(t)P(t_{0}|t)\right] , & \label{FP-epsilon} 
\end{eqnarray}
 where the external force has introduced the additional drift \( \underline{\epsilon }(t) \).
\textbf{}According to linear response theory \cite{kubo}, the response \( R_{\alpha _{i}\alpha _{j}}(t_{0},t) \)
of the system is then given through the functional derivative 

\begin{equation}
\label{R(t,s)}
R_{\alpha _{i}\alpha _{j}}(t_{0},t)=\frac{\delta \left\langle \alpha _{i}(t)\right\rangle }{\delta \epsilon _{j}(t_{0})}.
\end{equation}
 Notice that, due to the time dependence of the coefficients \( B_{ij}(t_{0};t) \),
the response function will contain modifications to the Markovian response. 

The response function can be calculated from the equation for the first moment
\( \left\langle \alpha _{i}(t)\right\rangle  \) which follows from \textbf{}the
Fokker-Planck equation (\ref{FP-epsilon}). We obtain \textbf{}

\textbf{
\begin{equation}
\label{F-D-linear-response}
R_{\alpha _{i}\alpha _{0j}}(t_{0},t)=\frac{m}{k_{B}}\frac{1}{T(t)}\frac{\partial }{\partial t_{0}}C_{\alpha _{i}\alpha _{0j}}(t_{0},t),
\end{equation}
}which constitutes the formulation of the fluctuation-dissipation theorem. This
expression indicates that the fluctuation-dissipation theorem remains valid
when the system is in local quasi-equilibrium; the temperature entering that
expression is that of the system at local quasi-equilibrium. If one uses Eqs.
(\ref{Tfuncion deATo}) and \textbf{}(\ref{F-D-linear-response}), it can also
be expressed as \textbf{
\begin{equation}
\label{FDLinear 2}
R_{\alpha _{i}\alpha _{0j}}(t_{0},t)=\frac{m}{k_{B}}\frac{A^{-1}}{T_{B}(t)}\frac{\partial }{\partial t_{0}}C_{\alpha _{i}\alpha _{0j}}(t_{0},t).
\end{equation}
} The result (\ref{FDLinear 2}) can in general be interpreted as a violation
of the fluctuation-dissipation theorem \cite{kurchan9} and \cite{kurchan}.
When analyzing its equivalent form (\ref{F-D-linear-response}) it expresses
validity of the fluctuation-dissipation theorem in terms of the local quasi-equilibrium
temperature.

The quantity \( A \) has been derived in Section \textbf{II} from thermodynamic
grounds. An expression in terms of correlation functions can be obtained from
the equal-time correlation function \( \Phi _{\alpha \alpha }(t)\equiv \frac{1}{2}\int \underline{\alpha }\cdot \underline{\alpha }P(\underline{\alpha }_{0},t_{0})P(\underline{\alpha }_{0},t_{0}|\underline{\alpha },t)d\underline{\alpha }_{0}d\underline{\alpha } \).
After substituting Eq. (\ref{fokker-planck general}) and integrating by parts,
one obtains the evolution equation 

\begin{equation}
\label{evol. equal time corr}
\frac{d}{dt}\Phi _{\alpha \alpha }(t)=-B(t_{0};t)\left\langle \underline{X}(t)\cdot \underline{\alpha }(t)\right\rangle +\frac{k_{B}T(t)}{m}B(t_{0};t),
\end{equation}
 where for simplicity, we have considered \( \underline{\underline{B}}(t_{0};t)=B(t_{0};t)\underline{\underline{1}} \),
with \( \underline{\underline{1}} \) the unit tensor. For times \( t\gg \beta ^{-1}(t_{0};t) \),
the time derivative of Eq. (\ref{evol. equal time corr}) may be neglected and
the system relaxes to the reference state, leading to the expression

\begin{equation}
\label{A}
A\equiv \frac{m}{k_{B}T_{B}(t)}\left\langle \underline{X}(t)\cdot \underline{\alpha }(t)\right\rangle _{qe},
\end{equation}
 where \( \left\langle ...\right\rangle _{qe} \) represents the average in
respect to the local quasi-equilibrium probability density \( P_{qe}(t) \).
\textbf{}This equation indicates that \( A \) may in general \textbf{}be expressed
by means of the moments of the probability density. \textbf{}According to Eqs.
(\ref{Tfuncion deATo}) and (\ref{A}), it follows that the temperature \( T(t) \)
entering the Gibbs equation (\ref{Gibbs ec}) results from the contribution
of the energies related to the corresponding degrees of freedom \( \underline{\alpha } \)
by 

\begin{equation}
\label{def. T por ETCF}
T(t)\equiv \frac{m}{k_{B}}\left\langle \underline{X}(t)\cdot \underline{\alpha }(t)\right\rangle _{qe}.
\end{equation}
An important fact to be noticed from this expression is that the temperature
is not robust since it implicitly depends on the number of relevant variables
or observables used to describe the system, as has been evidenced in \cite{origin}.
\textbf{}Given that the probability density of the reference state depends on
the initial condition, it follows that \( T(t) \) is also a function of \( t_{0} \).
In the case when the force is linear, Eq. (\ref{def. T por ETCF}) reduces to
\( T(t)\cong \frac{m}{k_{B}}\left\langle \alpha ^{2}(t)\right\rangle _{qe} \),
which coincides with the corresponding result \textbf{}obtained when local thermodynami\textbf{c}
equilibrium between system and bath exists. In that case, when \( \underline{\alpha }=\vec{u} \)
with \( \vec{u} \) the velocity of a particle, Eq. (\ref{def. T por ETCF})
becomes the kinetic definition of the temperature.

\section{Slow dynamics in glassy systems}

As a first application of the theory \textbf{}presented, \textbf{}we will analyze
the dynamics of systems characterized by two sets of variables: fast \( \beta  \)-variables
and slow \( \alpha  \)-variables. \textbf{}The \( \beta  \)-variables relax
to a local equilibrium state characterized by the temperature of the bath \textbf{\( T_{B} \)},
whereas the \( \alpha  \)-variables relax to a quasi-equilibrium state characterized
by the temperature \( T(t) \). 

The probability density is now a function of the state vector \( \underline{\Gamma }=(\underline{\alpha },\underline{\beta }) \).
Since \( \beta  \)-variables equilibrate with the bath, we may assume that
the probability density \textbf{}characterizing the reference state can be factorized
in the form \( P_{qe}(t)=P_{le}(\underline{\beta })P_{qe}(\underline{\alpha },t) \),
with \( P_{le}(\underline{\beta }) \) and \( P_{qe}(\underline{\alpha },t) \)
stationary and non-stationary distributions. Consequently, \( s_{qe}(t)=s^{\beta }_{le}+s^{\alpha }_{qe}(t) \),
where \textbf{\( T_{B}\, \delta s^{\beta }_{le} \)} and \textbf{\( T(t)\, \delta s^{\alpha }_{qe} \)}
enter Gibbs equations similar to Eq. (\ref{Gibbs ec}). Consistent with the
analysis of Sec. \textbf{II}, the probability densities of the reference states
are given by \textbf{}
\begin{equation}
\label{P0 alfas betas}
P_{le}(\underline{\beta })=e^{\frac{m}{k_{B}\, T_{B}}[\mu ^{\beta }_{le}-e_{le}(\underline{\beta })]}\; \; \; \; \textrm{and}\; \; \; \; P_{qe}(\underline{\alpha },t)=P_{e}e^{\frac{m}{k_{B}\, T(t)}[\mu ^{\alpha }_{qe}-e_{qe}(\underline{\alpha })]}.
\end{equation}

By taking the time derivative of Eq. (\ref{p. gibbs}) and using Eq. (\ref{P0 alfas betas})
and the corresponding continuity equation, the entropy production can be written
as

\begin{equation}
\label{sigme glass}
\sigma =-\int P(t_{0}|t)\underline{v}(\underline{\Gamma },t)\cdot \frac{\partial }{\partial \underline{\Gamma }}\Theta (t_{0};t)d\underline{\Gamma },
\end{equation}
 where we have defined the nonequilibrium Massieu function \textbf{}

\textbf{
\begin{equation}
\label{Massieu}
\Theta (t_{0};t)\equiv \frac{k_{B}}{m}\ln P(t_{0}|t)+\frac{e_{le}(\underline{\beta })}{T_{B}}+\frac{e_{qe}(\underline{\alpha })}{T(t)},
\end{equation}
} which is the appropriate function in the entropy representation \cite{callen},
\cite{N-part}.

Since \( \underline{v}(\underline{\Gamma },t)\equiv \left( \underline{v}_{\beta },\underline{v}_{\alpha }\right)  \)
and \( \frac{\partial }{\partial \underline{\Gamma }}\equiv \left( \frac{\partial }{\partial \underline{\beta }},\frac{\partial }{\partial \underline{\alpha }}\right)  \),
from the entropy production (\ref{sigme glass}) we can \textbf{}obtain the
expressions for \( \underline{v}_{\beta } \) and \( \underline{v}_{\alpha } \)
(see Eq. (\ref{linear laws 2})). In general, the rules of nonequilibrium thermodynamics
introduce couplings between the velocities \( \underline{v}_{\beta } \) and
\( \underline{v}_{\alpha } \) and the corresponding forces \textbf{}through
the mobility \( \underline{\underline{B}} \) which is in general a function
of \( t_{0} \), \( t \), \( \underline{\alpha } \) and \( \underline{\beta } \). 

In this case, since no couplings appear due to the separation of time scales,
the mobility matrix \( \underline{\underline{B}} \) can be written as the direct
sum of two submatrices \( \underline{\underline{B}}^{\alpha } \) and \( \underline{\underline{B}}^{\beta } \).
For simplicity, we will assume that the \( B_{ij}^{\beta } \)'s are constant
since the \( \beta  \)-variables decay to a local equilibrium state \cite{degroot}.
The \( B_{ij}^{\alpha } \)'s will be functions of two times, \( B_{ij}^{\alpha }=B_{ij}^{\alpha }(t_{0},t) \). 

Now, by taking into account that \( \beta  \)-variables thermalize at \( T_{B} \)
and \( \alpha  \)-variables at \( T(t) \), \cite{tartaglia}-\cite{theo},
after substituting the expressions for \( \underline{v}_{\beta } \) and \( \underline{v}_{\alpha } \)
into the continuity equation one obtains the generalized Fokker-Planck equation 

\textbf{
\begin{eqnarray}
\frac{\partial P(t_{0}|t)}{\partial t}=\frac{\partial }{\partial \underline{\alpha }}\cdot \underline{\underline{B}}^{\alpha }(t_{0};t)\left[ \underline{X}^{\alpha }(\underline{\alpha })P(t_{0}|t)+\frac{k_{B}T(t)}{m}\frac{\partial P(t_{0}|t)}{\partial \underline{\alpha }}\right] +\; \; \; \; \; \; \; \; \; \; \; \; \; \; \; \; \;  &  & \nonumber \\
\frac{\partial }{\partial \underline{\beta }}\cdot \underline{\underline{B}}^{\beta }\left[ \underline{X}^{\beta }(\underline{\beta })P(t_{0}|t)+\frac{k_{B}T_{B}}{m}\frac{\partial P(t_{0}|t)}{\partial \underline{\beta }}\right] , & \label{FP glass} 
\end{eqnarray}
} where, according with Eqs. (\ref{X force}) and (\ref{P0 alfas betas}), \( \underline{X}^{\beta }(\underline{\beta })=\frac{\partial e_{le}(\underline{\beta })}{\partial \underline{\beta }} \)
and \( \underline{X}^{\alpha }(\underline{\alpha })=\frac{\partial e_{qe}(\underline{\alpha })}{\partial \underline{\alpha }} \). 

One may obtain the Fokker-Planck equation for the slow \( \alpha  \)-variables
by multiplying Eq. (\ref{FP glass}) by \( P(\underline{\Gamma }_{0},t_{0}) \)
and integrating over \( \underline{\beta }_{0} \) and \( \underline{\beta } \)
taking into account that the current \( P(t_{0}|t)\underline{v}_{\beta } \)
vanishes at the boundaries. Our result is Eq. (\ref{fokker-planck general}).
In similar form, the evolution equation for the \( \beta  \)-variables can
be derived by integrating over \( \underline{\alpha }_{0} \) and \( \underline{\alpha } \).
The resulting Fokker-Planck equation is \textbf{}

\begin{equation}
\label{FP betas glass}
\frac{\partial P(\underline{\beta },t)}{\partial t}=\frac{\partial }{\partial \underline{\beta }}\cdot \underline{\underline{B}}^{\beta }\cdot \left[ \underline{X}^{\beta }(\underline{\beta })P(\underline{\beta },t)+\frac{k_{B}T_{B}}{m}\frac{\partial }{\partial \underline{\beta }}P(\underline{\beta },t)\right] ,
\end{equation}
where \( P(\underline{\beta },t) \) is the single-time distribution function. 

The computation of the moments of the probability density following the procedure
indicated in Sec. \textbf{III}, allows one to find the statistical expression
of the temperatures 

\begin{equation}
\label{T glass}
T_{B}\equiv \frac{m}{k_{B}}\left\langle \underline{X}^{\beta }(\underline{\beta })\cdot \underline{\beta }\right\rangle _{le},
\end{equation}
 and

\begin{equation}
\label{Talfa-glass}
T(t)\equiv \frac{m}{k_{B}}\left\langle \underline{X}^{\alpha }(\underline{\alpha },t)\cdot \underline{\alpha }\right\rangle _{qe}.
\end{equation}
 Eq. (\ref{T glass}) implies that the fast degrees of freedom satisfy the equipartition
law and its energy is a quadratic function in \( \beta  \)'s, whereas (\ref{Talfa-glass})
is a definition of the temperature of the system in terms of the correlations
at the quasi-equilibrium state.

>From Eq. (\ref{FP glass}), we may compute the response of the system to an
external perturbation and derive \textbf{}the fluctuation-dissipation theorem
\textbf{}related to both sets of variables. Following the procedure indicated
in Sec. \textbf{III}, the response functions of \textbf{}the fast variables
\textbf{}satisfy the relation \textbf{}

\begin{equation}
\label{FD betas }
R_{\beta _{i}\beta _{0j}}(t_{0},t)=\frac{m}{k_{B}}\frac{1}{T_{B}}\frac{\partial }{\partial t_{0}}C_{\beta _{i}\beta _{0j}}(t_{0},t),
\end{equation}
 where \( R_{\beta _{i}\beta _{0j}}(t_{0},t)=\frac{\delta \left\langle \beta _{i}(t)\right\rangle }{\delta \epsilon _{j}(t_{0})} \).
The corresponding expression for the response \textbf{}of the \( \alpha  \)-variables
is given in Eq. (\ref{F-D-linear-response}). From Eq. (\ref{FD betas }) it
follows that the fast degrees of freedom satisfy the fluctuation-dissipation
theorem in terms of the temperature of the bath. \textbf{}

\section{Brownian motion in a granular gas }

As a second application of our formalism, we will discuss in this section the
Brownian motion of a grain in a granular gas, which has been previously addressed
in \cite{dufty1} by means of kinetic theory for systems with inelastic collisions.
\textbf{}The phase-space vector is now \( \underline{\alpha }=(\vec{u},\vec{r}) \),
where \( \vec{u} \) and \( \vec{r} \) are \textbf{}the \textbf{}velocity and
position of the Brownian grain. The corresponding continuity equation for \( P(t_{0}|t) \)
is given by 

\begin{equation}
\label{continuidad granular}
\frac{\partial P(t_{0}|t)}{\partial t}+\nabla \cdot \vec{u}P(t_{0}|t)=-\frac{\partial }{\partial \vec{u}}\cdot \left[ P(t_{0}|t)\vec{v}_{\vec{u}}(t_{0};t)\right] ,
\end{equation}

We will assume that the bath is in a non-stationary homogeneous cooling state
characterized by the kinetic temperature \( T_{B}(t) \), \cite{dufty1}. The
kinetic temperature is given by \textbf{\( T(t)=\frac{m}{k_{B}}\int \frac{u^{2}}{2}P(\vec{u}_{0},t_{0})P(\vec{u}_{0},t_{0}|\vec{u},t)d\vec{u}_{0}d\vec{u} \),}
and has to be understood as the temperature of a \textbf{}dilute gas of Brownian
particles suspended in the granular gas. According to Eq. (\ref{P0 a la entropic}),
in this case the local quasi-equilibrium probability density can be taken as 

\textbf{
\begin{equation}
\label{Po granular}
P_{qe}(t)=e^{\frac{m}{k_{B}T(t)}[\mu _{qe}-\frac{1}{2}\vec{u}^{2}]},
\end{equation}
}where \( \mu _{qe} \) normalizes \( P_{qe}(t) \). 

Once the \( P_{qe}(t) \) has been specified, the Fokker-Planck equation for
the Brownian gas may be derived by using Eqs. (\ref{p. gibbs}), (\ref{continuidad granular}),
(\ref{Po granular}). Following \textbf{}the steps indicated in Sec. \textbf{II},
we finally obtain the generalized Fokker-Planck equation

\begin{equation}
\label{FP-Granujar gases}
\frac{\partial P(t_{0}|t)}{\partial t}+\nabla \cdot \vec{u}P=\frac{\partial }{\partial \vec{u}}\cdot B(t_{0},t)\left[ \vec{u}P(t_{0}|t)+\frac{k_{B}T_{B}(t)}{m}A\frac{\partial P(t_{0}|t)}{\partial \vec{u}}\right] ,
\end{equation}
 where we have assumed locality in \( \vec{u} \)-space for which the velocity
\( \vec{v}_{\vec{u}}(t_{0};t) \) couples only to the gradient of the chemical
potential at the same value of \( \vec{u} \) in accordance with Eq. (\ref{produccion S}).
\textbf{}By considering the action of an external force field \( \vec{\epsilon }(t) \)
which can be an impulsive one (tapping force), we may derive by following the
steps indicated in Sec. \textbf{III}, the fluctuation-dissipation relation 
\begin{equation}
\label{FDT-velocidad}
R_{\vec{u}\vec{u}_{0}}(t_{0},t)=\frac{m}{k_{B}}\frac{A^{-1}}{T_{B}(t)}\frac{\partial }{\partial t_{0}}C_{\vec{u}\vec{u}_{0}}(t_{0},t).
\end{equation}
 Recent computer simulations of diffusion in dilute granular gases confirm the
validity of Eq. (\ref{FDT-velocidad}), \cite{loreto}. The corresponding integrated
version of the generalized fluctuation-dissipation theorem has been also obtained
in dense granular systems in \cite{Nature-kurchan}. \textbf{}Now, by comparing
Eq. (\ref{FP-Granujar gases}) with the corresponding one found in \cite{dufty1}
\textbf{}by following kinetic theory methods, we \textbf{}may identify

\begin{equation}
\label{identificaciones 1}
A=\frac{\left( 1+\lambda \right) }{2}\; \; \; ;\; \; \; B(t_{0},t)=\frac{1}{2}\left( 1+\lambda \right) \gamma _{e}(t_{0},t),
\end{equation}
 where the restitution coefficient \( \lambda (\leq 1) \) of the Brownian grains
has been assumed constant, and \( \gamma _{e}(t_{0},t) \) is the friction coefficient
for the case when elastic collisions take place, \cite{dufty1}. 

>From the Fokker-Planck equation (\ref{FP-Granujar gases}), it follows that
the evolution equation of the second moment of the velocity of the Brownian
grains can be expressed in the form

\begin{equation}
\label{evol T}
\frac{d}{dt}T(t)=-2B(t_{0};t)\left[ T(t)-AT_{B}(t)\right] .
\end{equation}
 which is a particular case of Eq. (\ref{evol. equal time corr}). Eq. (\ref{evol T})
establishes that the relation between system and bath temperatures at the local
quasi-equilibrium state given by Eq. (\ref{Tfuncion deATo}), is valid for times
\( t\gg B^{-1}(t_{0};t) \). 

The relation between \( T(t) \) and \( T_{B}(t) \) can be generalized in order
to describe shorter time scales by incorporating the thermal relaxation of the
heat bath through the evolution equation \( \frac{d}{dt}T_{B}(t)=-\zeta (t)T_{B}(t) \),
with \( \zeta (t) \) the cooling rate. Assuming that \textbf{\( 0<\zeta (t)<2B(t_{0};t) \)},
the time evolution of \( \frac{T(t)}{T_{B}(t)} \) is then given by \cite{dufty1}

\begin{equation}
\label{evol T/To}
\frac{d}{dt}\frac{T(t)}{T_{B}(t)}+2\left( 1-\xi \right) B(t_{0};t)\frac{T(t)}{T_{B}(t)}=2B(t_{0};t)A,
\end{equation}
 where we have defined \( \xi \equiv \frac{\zeta (t)}{2B(t_{0};t)} \). From
Eq. (\ref{evol T/To}), we then obtain 

\textbf{
\begin{equation}
\label{T short times}
T(t)=\frac{A}{\left( 1-\xi \right) }T_{B}(t),
\end{equation}
}which gives the relation between the system and bath temperatures for times
\( t\gg \left[ \left( 1-\xi \right) B(t_{0};t)\right] ^{-1} \). 

By contracting the description, we enter the diffusion regime described by the
density \( \rho (\vec{r}^{\prime },t_{0}|\vec{r},t)=m\int P(t_{0}|t)d\vec{u} \),
whose evolution is governed by a Smoluchowski equation. This equation may be
derived by approximating the hierarchy of evolution equations of the first three
moments of the probability density \( P(t_{0}|t) \), \cite{nosotros}. One
obtains, 

\begin{equation}
\label{Smoluchowski 1}
\frac{\partial }{\partial t}\rho (t_{0}|t)=D_{eff}(t_{0},t)\nabla ^{2}\rho (t_{0}|t),
\end{equation}
 where \( D_{eff}(t_{0},t) \) is the effective diffusion coefficient given
by

\begin{equation}
\label{Deff granular 2}
D_{eff}(t_{0},t)\equiv \frac{k_{B}T_{B}(t)}{mB(t_{0},t)}\frac{A}{\left( 1-\xi \right) ^{2}},
\end{equation}
 which has been obtained by taking into account Eqs. (\ref{evol T/To}) and
(\ref{T short times}). Eq. (\ref{Deff granular 2}) was derived in \cite{dufty1}
by means of kinetic theory. In particular, when the cooling rate of the bath
satisfies \( \zeta (t)\ll 2B \), the expression of the diffusion coefficient
(\ref{Deff granular 2}) reduces to \( D_{eff}(t_{0},t)\equiv \frac{k_{B}T_{B}}{mB(t_{0},t)}A \).
In the general case, the dynamics of the system may be described \textbf{}by
means of the corresponding Fokker-Planck equation in which the probability density
of the reference state has the form \( P_{qe}(t)=e^{\frac{m}{k_{B}T_{B}(t)}A^{-1}\left( 1-\xi \right) [\mu _{qe}-\frac{1}{2}\vec{u}^{2}]} \),
and the coefficients \( A \) and \( B(t_{0},t) \) satisfy more general relations
than Eq. (\ref{identificaciones 1}).

The Fokker-Planck and Smoluchowski equations derived previously may also describe
diffusion \textbf{}in a colloidal system. \textbf{}In that case, the diffusion
coefficient (\ref{Deff granular 2}) accounts for the violation of the Stokes-Einstein
relation reported in dense granular systems \cite{Nature-kurchan} and supercooled
colloidal fluids \cite{coll-glasses}.

\section{Discussion}

In this paper, we have proposed an extension of Onsager's theory, originally
formulated to describe \emph{nonequilibrium aged} systems, \textbf{}in order
to analyze slow relaxation dynamics. Our theory is based on the assumption of
the existence of a local quasi-equilibrium state in phase space \textbf{}characterized
by a non-stationary probability density. This hypothesis entails the existence
of a time-dependent temperature of the system \( T(t) \) reflecting \textbf{}lack
of thermal equilibrium between system and bath, which enables us to formulate
a Gibbs \textbf{}equation. \textbf{}The relation between the local quasi-equilibrium
and bath temperatures was given in terms of a scale function obtained by taking
into account the influence of the bath. 

The entropy production \textbf{}derived from the Gibbs entropy postulate leads
to the explicit expression of the probability current in phase space and, consequently,
to the generalized Fokker-Planck equation (\ref{fokker-planck general}). Memory
effects were incorporated in the description through the time dependence of
the coefficients entering into that equation. They are responsible for the dependence
of the characteristic \textbf{}relaxation times on the initial (waiting) time.
Our formalism could straightforwardly be extended to the case of a description
in terms of fields \textbf{\( \underline{\alpha }(\vec{k}) \)}, \cite{hydrofluc},
\cite{supercooled}.

We have shown that the fluctuation-dissipation relation remains valid \textbf{}in
terms of the local quasi-equilibrium temperature \( T(t) \) of the system whose
\textbf{}value depends on the time scale considered. \textbf{}The fact that
the local quasi-equilibrium temperature depends on time and on the number of
variables relevant to characterize the state, clarifies why two systems with
slow relaxation dynamics put in thermal contact do not necessarily satisfy a
``zero law''.

As a first application, we have analyzed the slow dynamics of glasses. We have
shown that the dynamics of the system can be described by a generalized Fokker-Planck
equation governing the time evolution of both the fast and slow degrees of freedom.
By using this equation, we have verified that the fluctuation-dissipation theorem
(\ref{FD betas }) is satisfied by the fast modes. For the slow modes the theorem
(\ref{F-D-linear-response}) is valid in terms of the local quasi-equilibrium
temperature. 

We have also studied the problem of Brownian diffusion in a dilute granular
gas reproducing the Fokker-Planck and Smoluchowski \textbf{}equations derived
previously by means of kinetic theory of granular gases. These equations may
also be used to describe the dynamics of supercooled colloidal suspensions leading
to a generalization of the Stokes-Einstein relation. 

The theory presented offers a unified description of systems undergoing slow
relaxation and constitutes a generalization of the Onsager theory describing
the evolution of fluctuations around nonequilibrium \emph{aging} states.

\section{Acknowledgments}

We want to acknowledge Dr. D. Reguera for valious comments. I.S.H. acknowledges
UNAM-DGAPA for economic support. This work was partially supported by DGICYT
of the Spanish Government under Grant No. PB2002-01267.

\end{document}